\documentclass[a4paper]{jpconf}
\usepackage{graphicx}
\begin{document}
\title{Heisenberg antiferromagnets with exchange and cubic anisotropies}

\author{G Bannasch$^1$ and W Selke$^2$}
\address{$^1$ MPI f\"ur Physik komplexer Systeme, 01187 Dresden, Germany}  
\address{$^2$ Institut f\"ur Theoretische Physik, RWTH Aachen University
  and JARA-SIM, 52056 Aachen, Germany}

\ead{selke@physik.rwth-aachen.de}

\begin{abstract}
We study classical Heisenberg antiferromagnets with uniaxial exchange
anisotropy and a cubic anisotropy term on simple cubic
lattices in an external magnetic field using ground state considerations
and extensive Monte Carlo simulations. In addition to the
antiferromagnetic phase field--induced spin--flop
and non--collinear, biconical phases may occur. Phase diagrams and
critical as well as multicritical phenomena are discussed. Results
are compared to previous findings.
\end{abstract}

\section{Introduction}
Recently, there has been a renewed interest in uniaxially
anisotropic Heisenberg antiferromagnets in a field, for
many years known to display antiferromagnetic and field--induced
spin--flop phases. This interest
is due to various reasons, among others, (i) to clarify
the phase diagram of the prototypical XXZ model, especially
for the square lattice \cite{holt1,landau,holt2}, (ii) to study
multicritical, like bi-- and tetracritical, behavior
 \cite{vic,folk}, and (iii) to elucidate ground states as well as
thermal properties of low--dimensional quantum magnets exhibiting, possibly,  
'supersolid' magnetic (i.e. non--collinear 'biconical' \cite{fisher})
 structures \cite{lafl,seng,peters}. Of
course, rather recent pertinent experiments also should be
mentioned \cite{buch,bog}.

\section{Results}

As a starting point of theoretical studies on uniaxially anisotropic
Heisenberg antiferromagnets, one often considers the
XXZ model, with the Hamiltonian

\begin{equation}
  {\cal H}_{\mathrm{XXZ}} = J \sum\limits_{i,j}
  \left[ \, \Delta (S_i^x S_j^x + S_i^y S_j^y) + S_i^z S_j^z \, \right]
  \; - \; H \sum\limits_{i} S_i^z
\end{equation}

\noindent
where $J > 0$ is the exchange coupling between spins
being located on neighboring lattice sites $i$ and $j$. $\Delta$
is the exchange anisotropy, $1 > \Delta > 0$, and $H$ is the applied
magnetic field along the easy axis, the $z$--axis. The model has
been found to display in the (temperature T, field H)--plane
antiferromagnetic (AF) and spin--flop (SF) phases on square and simple cubic
lattices. In two dimensions, the transition between the two
phases had been argued to be of first order in case of the spin--1/2  
quantum case \cite{holt2,troy}. In contrast, in the classical
case (with spin vectors of length one) there appears to be a narrow
disordered phase in between the two ordered phases. The
three phases seem to meet at zero temperature
in a 'hidden tetracritical point' \cite{landau,holt2}
at the critical field $H_c$. That point is a highly
degenerate ground state, where also
biconical (BC) spin configurations are stable \cite{holt2}. The 
antiferromagnetic, biconical, and spin--flop classical spin
configurations are shown in Fig. 1. For the XXZ model on the
simple cubic lattice, the
phase diagram has been believed to show the same topology as in
the mean--field approximation, with a direct transition of first
order between the AF and SF phases, ending in a bicritical point at
which the two critical phase boundary lines between the paramagnetic (P)
phase and the AF as well as the SF phases meet with the AF-SF transition
line \cite{binder}. Recently, this scenario has been scrutinized
using renormalization group methods \cite{vic,folk}. One of the aims
of our study is to shed light upon this issue, applying extensive 
Monte Carlo simulations.

\begin{figure}[h]
\includegraphics[width=16pc]{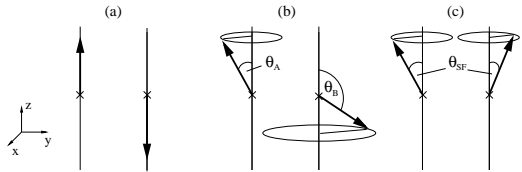}\hspace{2pc}%
\begin{minipage}[b]{16pc}\caption{\label{label}Spin orientations on
    neighboring sites showing antferromagnetic (a), biconical
    (b), and spin--flop (c) ground state structures in the XXZ model.}
\end{minipage}
\end{figure}

The phase diagram obtained from our simulations,
analyzing systems with up to $32^3$ spins in runs of, at
least, $10^7$ Monte Carlo steps per spin, is shown
in Fig. 2. Indeed, setting $\Delta= 0.8$, we can
locate the triple
point of the  AF-P, SF-P, and AF-SP boundary lines accurately, at
$k_BT_t/J = 1.025 \pm 0.015$ and $H_t/J= 3.90 \pm 0.03$, differing quite
substantially from the old estimate based on appreciably shorter
simulations \cite{binder}.

\begin{figure}[h]
\includegraphics[width=17pc]{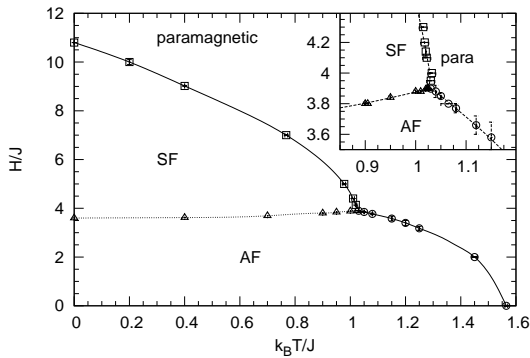}
\begin{minipage}[b] {17pc}
\caption{\label{label}Phase diagram of the XXZ
  antiferromagnet, with $\Delta= 0.8$. Inset: Vicinity of the triple point.}
\end{minipage}\hspace{2pc}%
\end{figure}

Moreover, studying the transitions in
the vicinity of that triple point, we do not observe deviations
from the previously anticipated bicritical scenario \cite{binder}, with 
the AF-P boundary line belonging to the Ising, and the
SF-P boundary line belonging to the XY universality
class. We identify the universality classes from determining
critical exponents, e.g., of the (staggered) longitudinal
and tranverse susceptibilities, and from determining the critical Binder
cumulants of the AF and SF order parameters \cite{ban2}. Note that in a
renormalization group calculation 
to high loop order, a  different scenario had been put forward
\cite{vic}, favouring the existence of a tetracritical
point or some kind of critical end point. In that scenario, one
expects either a stable BC phase near ($T_t,H_t$) or
at least one of the AF-P and SF-P boundaries should become
a line of transitions of first order near
that point \cite{vic}. In our simulations, we observe that
biconical spin configurations show up close to the AF-SF phase
boundary at low temperatures, reflecting the, again, high 
degeneracy of the ground state at the critical field 
separating the AF and SF structures. However, these
configurations do not destroy, in contrast to the situation
in two dimensions, the direct AF-SF transition of 
first order \cite{ban}. Actually, applying also renormalization group
arguments, it has been suggested very recently that
the type of the triple point may depend on the strength of the
anisotropy, allowing for a bicritical point \cite{folk}. Perhaps, Monte Carlo
studies on different values of the anisotropy may provide
further insights.    
 
Adding, especially, single--ion terms due to crystal--field
anisotropies to the XXZ model, eq. (1), BC spin configurations 
may be stabilized over a range of fields in the ground state. This
behavior has been observed for quantum and
classical spins on chains and square
lattices for a quadratic single--ion
term \cite{lafl,seng,peters,holt3}. Here we shall add to 
${\cal H}_{\mathrm{XXZ}}$ a cubic anisotropy term of the
form \cite{aha}

\begin{equation}
  {\cal H}_{\mathrm{CA}} =  F \sum\limits_{i} \left[(S_i^x)^4 + (S_i^y)^4+
(S_i^z)^4 \right]
\end{equation}

\noindent
where $F$ denotes the strength of the cubic anisotropy. The sign of
$F$ determines whether the spins tend to align along the cubic
axes, for $F < 0$, case 1, or, for $F > 0$, case 2, in the
diagonal directions of the lattice. Because
of these tendencies, the BC, (i.e. BC1 or BC2), structures, as well as the
SF structures, show no full rotational invariance
in the $xy$--plane, in contrast
to the XXZ case discussed above. Now, obviously, the
discretized spin projections in the $xy$--plane
favour four directions \cite{ban2,ban}.

The resulting ground state phase diagram of the
full Hamiltonian, ${\cal H}_{\mathrm{XXZ}} + {\cal H}_{\mathrm{CA}}$ with
fixed exchange anisotropy, $\Delta =0.8$, and varying cubic term, $F$,  
may be determined numerically
without difficulty, as depicted in Fig. 3.

\begin{figure}[h]
\includegraphics[width=17pc]{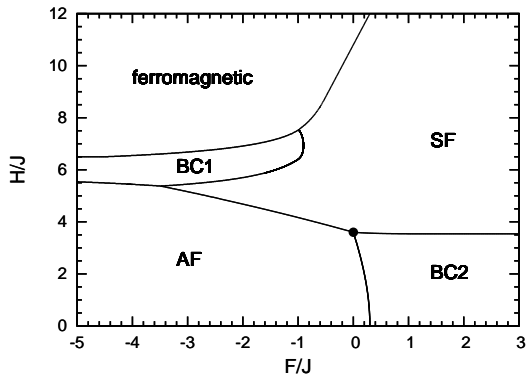}\hspace{2pc}%
\begin{minipage}[b]{17pc}\caption{\label{label}Ground states of the
    full Hamiltonian with exchange anisotropy, $\Delta= 0.8$, and
    the cubic term.}
\end{minipage}
\end{figure}

Note that, for $F <0$, the transitions to the BC1 structures are typically
of first order, with a jump in the tilt angles, $\Theta$, with respect
to the $z$-axis, see Fig. 1, characterizing the BC
configurations. However, in the reentrance region between
the SF and BC1 structures at $F/J$ close to -1, the change in
the tilt angles seems to be smooth at the transitions \cite{ban2}.

Obviously, at non-zero temperatures, several interesting scenarios
leading, possibly, to multicritical behaviour, where
AF, SF, BC, and P phases meet, may exist. So far, we focussed
attention on two cases: (a) Positive cubic anisotropy $F > 0$, at
constant field $H/J = 1.8$ \cite{ban}, see Fig. 3. At small
values of $F$, there is an
AF ordering at low temperatures. Above a critical
value, $F_c =0.218...J$, the low--temperature phase is of BC2 type,
followed by the AF and P phases, when increasing the
temperature. The transition between the AF and P phases is found to
belong to the Ising universality class, while the transition between
the BC2 and AF phases seems to belong to the XY universality class,
with the cubic term being then an irrelevant
perturbation \cite{ban}. (b) Negative $F$, fixing the cubic
term, $F/J = -2$, and varying the field, see Fig. 3. In accordance with
the ground state analysis, we observe, at sufficiently low
temperature, $k_BT/J= 0.2$, AF, SF, BC1, and P phases, when increasing
the field, as depicted in Fig. 4. Interestingly, the BC1 phase
seems to become unstable when raising the temperature, with the other
phases being still present \cite{ban2}, see Fig. 5. This may suggest that the
three boundary lines between the BC1--P, SF--BC1, and
SF--P phases meet at a multicritical point. Of course, further
clarification and a search for other, possibly multicritical
scenarios at different strengths of the cubic term, $F$, are desirable.

\begin{figure}[h]
\begin{minipage}{14pc}
\includegraphics[width=14pc]{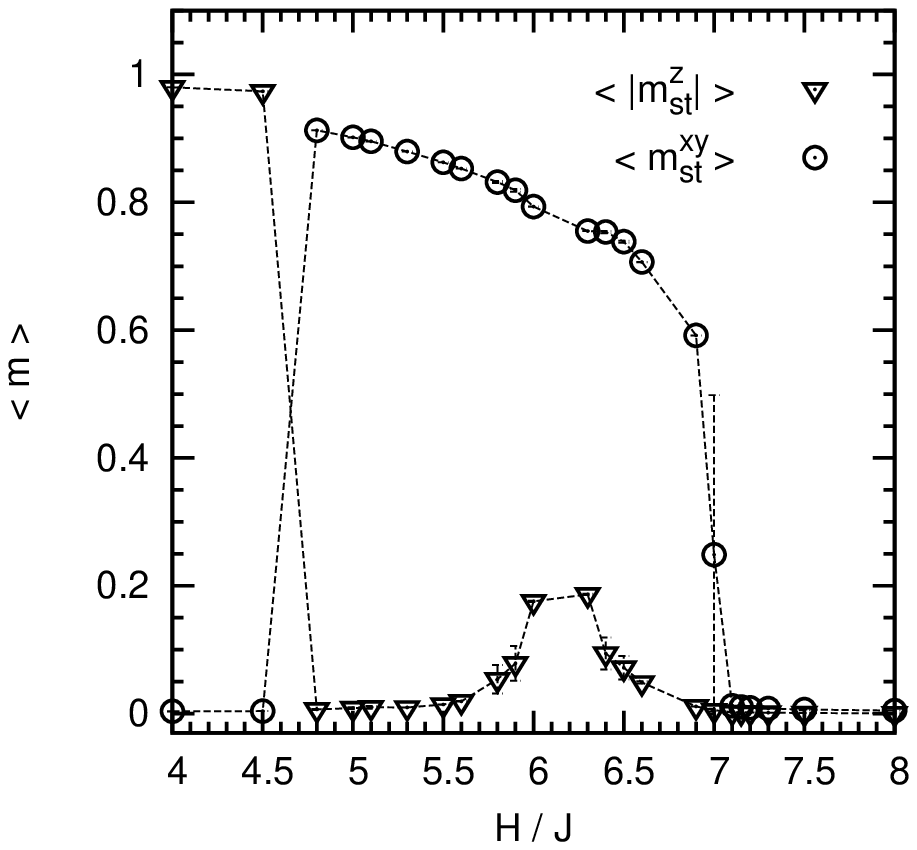}
\caption{\label{label} Staggered magnetizations versus field at $F/J=
  -2$ and $k_BT/J = 0.2$, indicating the AF, SF, BC1, and P phases,
  when increasing the field.}
\end{minipage}\hspace{2pc}%
\begin{minipage}{14pc}
\includegraphics[width=14pc]{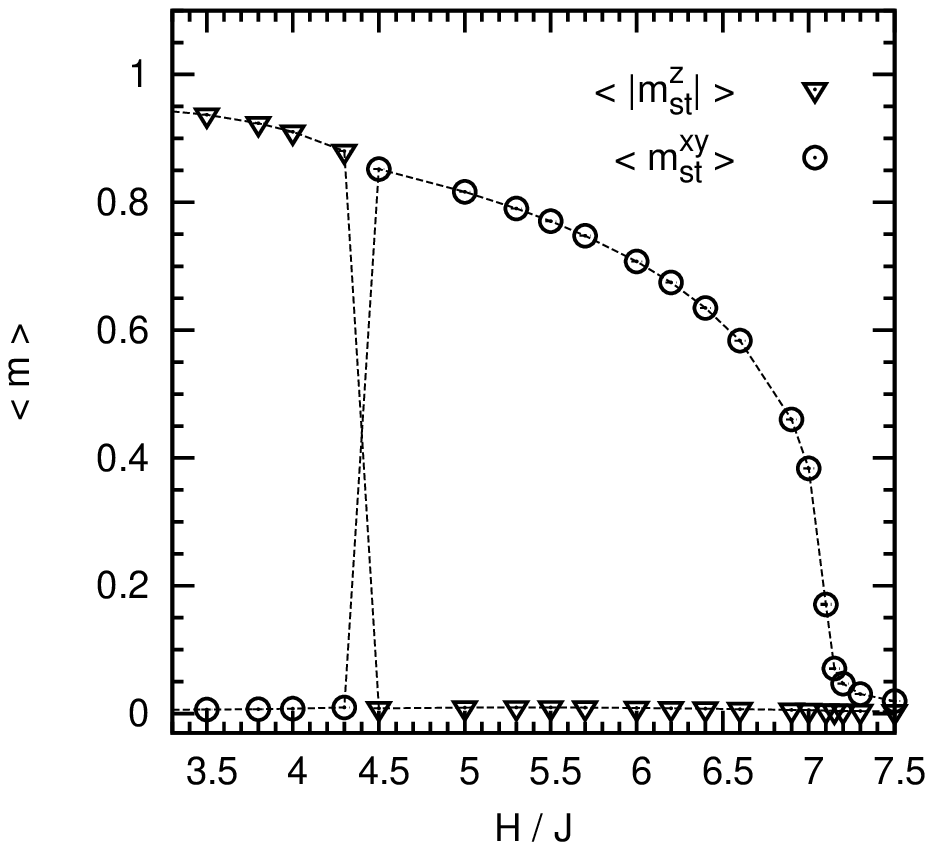}
\caption{\label{label}Staggered magnetizations versus
  field, at $F/J= -2$ and $k_BT/J= 0.4$, with the BC1 phase
  being squeezed out. Systems with $16^3$ spins are simulated.}
\end{minipage} 
\end{figure}

In summary, we have studied antiferromagnets with
fixed exchange and varying cubic anisotropies in a field on the 
simple cubic lattice. In the case of the XXZ model, the
nature of the triple point and the thermal role of BC structures 
have been clarified. By adding the cubic anisotropy, discretized
BC phases may be stabilized leading to interesting
critical and multicritical phenomena.        
  
\ack

We should like to thank A. N. Bogdanov, T.-C. Dinh, R. Folk,
M. Holtschneider, D. P. Landau, and D. Peters for useful discussions
and information.

\end{document}